\documentclass{PoS}
\usepackage{amsmath,amsbsy,subfig}

\newcommand{\mrm}{\mathrm}
\newcommand{\ovl}{\overline}
\newcommand{\eps}{\epsilon}

\newcommand{\ra}{\rightarrow}

\newcommand{\sWsq}{\sin^2\theta_\mathrm{W}}

\title{Non-Supersymmetric New Physics from M\o{}ller Scattering}

\ShortTitle{Non-SUSY New Physics from M\o{}ller Scattering}

\author{\speaker{Jackson Wu} \\
	Institute for Theoretical Physics, University of Bern\\
	Sidlerstrasse 5, 3012 Bern, Switzerland\\
        E-mail: \email{jbnwu@itp.unibe.ch}}

\abstract{We study in an effective operator approach how the effects of new physics from various scenarios containing extra $Z'$ gauge bosons or doubly charged scalars can affect, and thus be tested by the precision polarized Moeller scattering experiment. We give Wilson coefficients for various classes of models, and we deduce constraints on the parameter space of the relevant coupling constants and mixing angles from the results of the SLAC E158 experiment, and that projected from the future high precision JLAB experiment.}

\FullConference{European Physical Society Europhysics Conference on High Energy Physics\\
		 July 16 - 22, 2009\\
		 Krakow, Poland}

\begin{document}
\section{Introduction}
It is well-known that left-right asymmetry, $A_{LR}\equiv(d\sigma_L - d\sigma_R)/(d\sigma_L + d\sigma_R)$, provides avery clean probe of new physics. Currently, it has been measured by the SLAC E158 collaboration at the 10\% level~\cite{E158}:
\begin{equation}
A_{LR} = 131 \pm 14\,(\mrm{stat}) \pm 10\,(\mrm{syst})\,\mrm{ppb} \,.
\end{equation}
But when the 12~GeV upgrade at JLAB is completed, a 1~ppb measurement will be possible. At this precision,
polarized M\o{}ller scattering is sensitive to new physics (NP) such as extra $Z'$ gauge bosons ubiquitous in beyond the Standard Model (BSM) scenarios, and also the more exotic doubly charged scalars at the several TeV range. This will complement the LHC nicely, as it allows the extraction of coupling information of the new states once their masses are measured in the direct searches.

Another area M\o{}ller scattering can have an important impact is in probing the GeV scale hidden sector NP not accessible to high energy colliders. The existence of such a hidden sector is recently suggested by the results of the PAMELA and FERMI satellite obeservations in relation to dark matter~\cite{MPV09}, and M\o{}ller scattering can provide new insights for further dark model building.

\section{Effective Operators}
A powerful way to study NP above the Fermi scale is that of the effective operators, which allows their effects
to be parametrized systematically in a model-independent way. For polarized M\o{}ller scattering, the relevant effective Lagrangian is $\mathcal{L} = \mathcal{L}_{SM} + \mathcal{L}_6$, with
\begin{align}
-2\mathcal{L}_6 =
&\,\frac{c_{LL}}{\Lambda^2}(\bar{L}_a \gamma^\mu L_a)(\bar{L}_b \gamma_\mu L_b) + \frac{c_{LR}}{\Lambda^2}(\bar{L}_a \gamma^\mu L_a)(\ovl{e}_R \gamma_\mu e_R) + \notag \\
&\,\frac{c_{RR}}{\Lambda^2}(\bar{e}_R \gamma^\mu e_R)(\bar{e}_R \gamma_\mu e_R) +
\frac{d_{LL}}{\Lambda^2}(\bar{L}_a \gamma^\mu L_b)(\bar{L}_b \gamma_\mu L_a) + h.c. \,,
\end{align}
where $L=(\nu,e)^T_L$. 
%In particular, the relevant operators contributing to $A_{LR}$ are
%\begin{equation}
%\mathcal{O}_1 = \bar{e}_L \gamma^\mu e_L \, \bar{e}_L \gamma_\mu e_L \,, \quad
%\mathcal{O}_2 = \bar{e}_R \gamma^\mu e_R \, \bar{e}_R \gamma_\mu e_R 
%\end{equation}
For $\Lambda^2 \gg s \gg m_e^2$ valid for the 12~GeV experiment, one then has
\begin{align}
A_{LR} &\equiv A_{LR}^\mathrm{SM} + \delta A_{LR} 
= \frac{4G_\mu s}{\sqrt{2}\pi\alpha}\frac{y(1-y)}{1+y^4 +(1-y)^4} \left[\left(\frac{1}{4}-\sWsq^{\ovl{\,\mathrm{MS}}}\right)
+\frac{c^{\prime}_{LL}-c_{RR}}{4\sqrt{2}G_\mu\Lambda^2}\right] \,,
\end{align}
where $y = Q^2/s$, $s = 2m_e E_{beam}$, and $c_{LL}' \equiv c_{LL} + d_{LL}$. Here, $\delta A_{LR}$ denotes the deviation of $A_{LR}$ arising from NP from the SM prediction, and the one-loop SM radiative corrections are encoded in the precise value of $\sWsq$ evaluated in the $\ovl{\mathrm{MS}}$ scheme. Note that the renormalization group running effects between $\Lambda$ (typically a few TeV) and $\sqrt{s} \simeq 0.1$~GeV is about 10\% at most. Thus to first approximation, they can be neglected (see Ref.~\cite{CNW09} for more details).

We define here a useful variable
\begin{equation}
\delta_{LR}\equiv\frac{\delta A_{LR}}{A_{LR}^{SM}} = 
\frac{c_{LL}' - c_{RR}}{\sqrt{2}G_\mu(1-4\sWsq^{\ovl{\,\mathrm{MS}}})\Lambda^2} \,,
\end{equation}
which will be used extensively below to study NP effects in various BSM scenarios. As a benchmark for comparison, taking $Q^2 = 0.026$~GeV$^2$ and $y = 0.6$, $A_{LR}^{SM} = 1.47 \times 10^{-7}$, and one has $-0.337 < \delta_{LR} < 0.122$ at $95\%$ CL from the E158 measurement. In the JLAB case, if one assumes a total $1\sigma$ error of 1~ppb, one has a projected $95\%$ CL limit $|\delta_{LR}| < 0.0136$.
%$\sWsq^{\ovl{\,\mathrm{MS}}}= 0.23867$ in $Q^2 = 0$ limit~\cite{EM05}.

\section{New Physics Scenarios}
The non-supersymmetric BSM scenarios examined here fall into two categories. In one, NP arises at a high scale, which couples directly to the SM. The Wilson coefficients parametrizing the NP effects in $A_{LR}$ are summarized in Table~\ref{Tb:NPWil} for the high scale NP scenarios examined. In the other, NP can be of a low scale, and couplings to the SM happen only through mixing interactions. This is typical of models with a hidden sector.
\begin{table}[htbp]
\begin{center}
\begin{tabular}{cccccc}
New Physics & $c'_{LL}$ & $c_{RR}$ & $c'_{LL} - c_{RR}$ \\
\hline
$E_6$ GUT & $\frac{3\pi\alpha}{2c_W^2}(c_\beta+\sqrt{\frac{5}{27}}s_\beta)^2$ &
$\frac{3\pi\alpha}{2c_W^2}(\frac{1}{3}c_\beta-\sqrt{\frac{5}{27}}s_\beta)^2$ &
$\frac{4\pi\alpha}{3c_W^2}c_\beta(c_\beta + \sqrt{\frac{5}{3}}s_\beta)$ \\
$U(1)_R \times U(1)_{B-L}$ & $\frac{1}{4}\frac{g_Y^4}{g_R^2-g_Y^2}$ & $\frac{1}{4}\frac{(2g_Y^2-g_R^2)^2}{g_R^2 -g_Y^2}$ & $\frac{1}{4}(3g_{Y}^2-g_{R}^2)$ \\
$U(1)_{B-L}$ & $g^{\prime 2}$ & $g^{\prime 2}$ & 0 \\
$U(1)_X$ & $g^{\prime 2}z_L^2$ & $g^{\prime 2}z_e^2$ & $g^{\prime 2}(z_L^2 - z_e^2)$ \\
$P^{\pm\pm}$ (RH) & 0 & $-\frac{1}{2}|y_{ee}|^2$ & $\frac{1}{2}|y_{ee}|^2$ \\
$P^{\pm\pm}$ (LH) & $-\frac{1}{2}|y_{ee}|^2$ & 0 & $-\frac{1}{2}|y_{ee}|^2$ 
\end{tabular}
\end{center}
\caption{\label{Tb:NPWil} Summary of Wilson coefficients $c'_{LL}$ and $c_{RR}$ for various generic high scale NP models containing an extra $Z'$ gauge boson, or doubly charged scalars, $P^{\pm\pm}$.}
\end{table}

\subsection{Top-down $U(1)_X$ models}
Extra $U(1)$ gauge factors commonly arise when a high scale symmetry is broken down to produce the SM gauge group. This is exmeplified by $E_6$ GUT models, which has the symmetry breaking pattern
$E_6 \ra SO(10) \times U(1)_\psi \ra SU(5) \times U(1)_\chi \times U(1)_\psi \ra SM \times U(1)_\beta$, and the final extra neutral gauge boson, $X^\mu(\beta) = B^\mu_\psi\sin{\beta} + B^\mu_\chi\cos{\beta}$, a linear combination of the $\psi$ and $\chi$ gauge bosons. Another example come from left-right symmetric (LRS) models, which has the symmetry breaking pattern 
$SU(2)_L \times SU(2)_R \times U(1)_{B-L} \ra SU(2)_L \times U(1)_Y \times U(1)'$. 
\begin{figure}[htbp]
\centering
\subfloat[]{\label{fig:subf:a}
\includegraphics[width=2.7in]{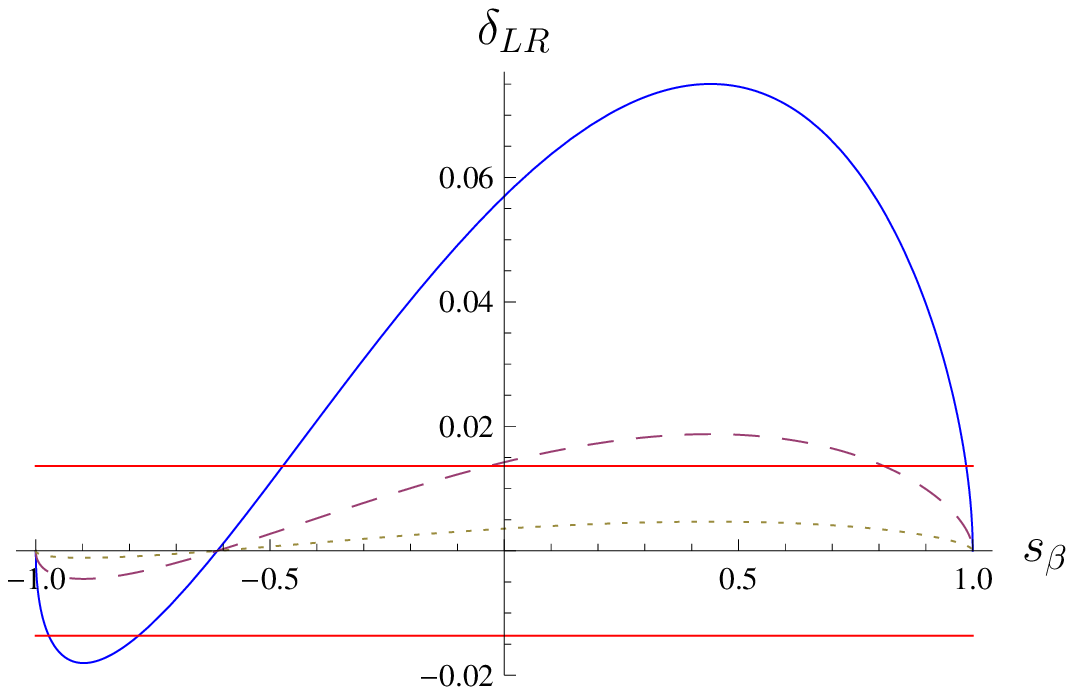}}
\hspace{0.2in}
\subfloat[]{\label{fig:subf:b}
\includegraphics[width=2.7in]{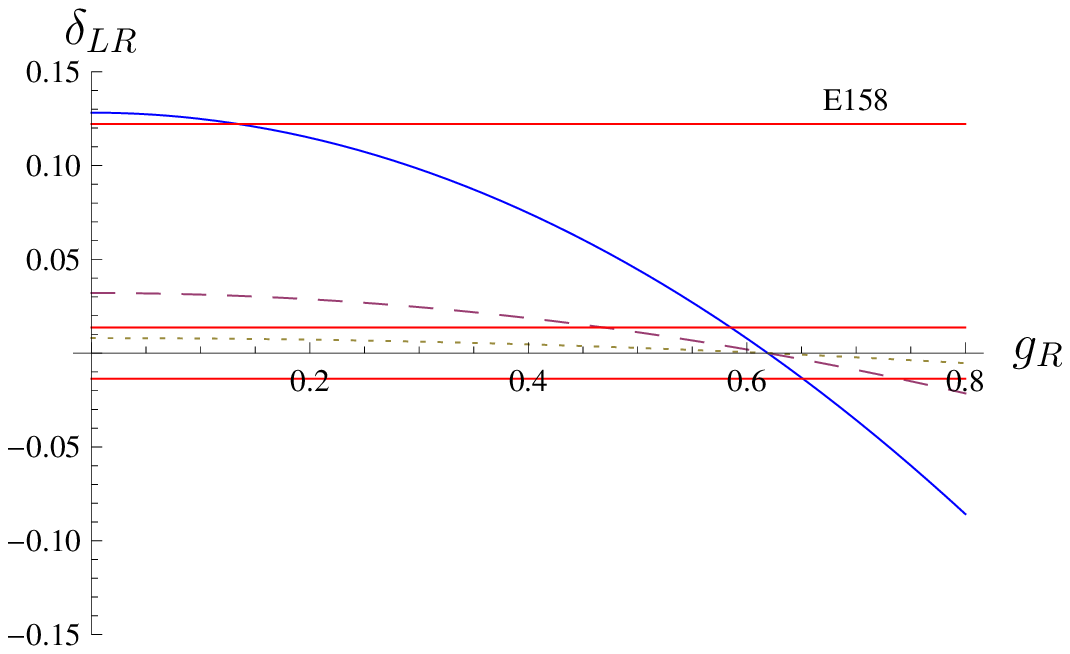}}
\caption{\label{fig:U1TD}
The ratio $\delta_{LR}$ in (a) GUT models and (b) LRS models. The solid, dashed and dotted lines denote $\Lambda = 1,\,2,\,4$~TeV. The unlabelled horizontal lines denote the projected JLAB limits.}
\end{figure}
The resulting $\delta_{LR}$ in the two scenarios are plotted in Fig.~\ref{fig:U1TD}. One can see that obtaining useful contraints on TeV scale NP depend very much on the higher precision of the future JLAB experiment.

\subsection{Bottom-up $U(1)_X$ Models}
There are phenomenologically interesting models where the extra $U(1)$ factors are not immediately related to a broken down larger symmetry. The assumption is that the SM~\footnote{We have introduced one sterile neutrino, $N_R$, per family to give the neutrinos masses.} is charged under the $U(1)_X$, which is known to be anomalous. Demanding that anomalies cancel then fixes these charges to be $z_Q = -z_L/3$, $z_u = 2z_L/3 - z_e$, $z_d = -4z_L/3 + z_e$, $z_N = 2z_L - z_e$, and $z_H = z_L - z_e$, where $z_f$ ($z_H$) denotes the fermion (Higgs) charge. In the special case where $U(1)_X$ is $U(1)_{B - L}$, $z_L = z_e$ and so 
$A_{LR} = A_{LR}^{SM}$.
\begin{figure}[htbp]
\begin{center}
\includegraphics[width=2.7in]{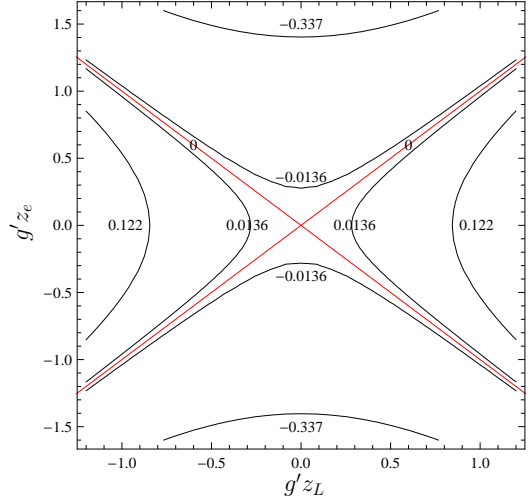}
\end{center}
\caption{\label{fig:U1BU} Contours of constant $\delta_{LR}$ for $\Lambda = 1$~TeV. The outer (inner) hyperbolas are the E158 (JLAB) limits. Other values of $\Lambda$ can be obtained via rescaling by a factor of $\Lambda/$TeV.}
\end{figure}
The allowed parameter space is shown by the contour plot of $\delta_{LR}$ in Fig.~\ref{fig:U1BU}. One can see that JLAB will be able to restrict the allowed parameter space significantly.

\subsection{Doubly Charged Scalars}
The doubly charged scalars are motivated by the Type II seesaw neutrino mass generation mechanisms, and they can be either $SU(2)_L$ doublets or singlets. Both type contribute to M\o{}ller scattering in the same way  (only their masses, $M_P$, and coupling to electrons, $y_{ee}$, matter) with appropriate Wilson coefficients. But crucially, the sign of their contribution is opposite, whose measurement can thus differentiate between different models of doubly charged scalars.
%\begin{figure}[htbp]
%\centering
%\includegraphics[width=3in]{dLRDCj.eps}
%\caption{Solid, dashed and dotted lines denote $M_P = 1,\,2,\,4$~TeV.}
%\end{figure}

\subsection{Hidden Sector/Shadow $Z'$ Models}
\begin{figure}[htbp]
\centering
\includegraphics[width=2.7in]{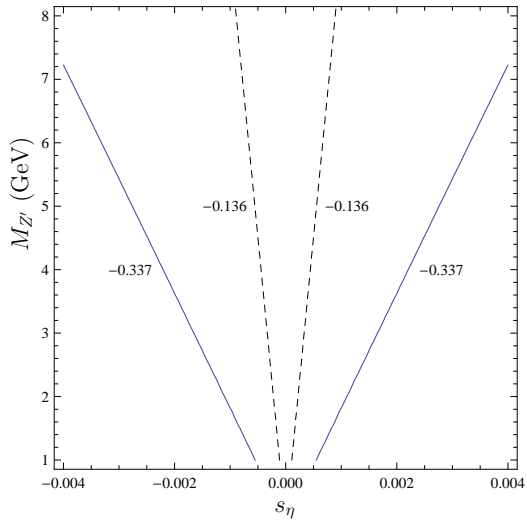}
\caption{\label{fig:hidZ} Contours of constant $\delta_{LR}$. Solid and dashed contours denote E158 and projected JLAB lower limits.}
\end{figure}

In this class of models, the extra $U(1)$ is a gauge symmetry of a hidden sector, which communicate with the SM only through gauge kinetic mixing. %$\frac{\eps}{2}X_{\mu\nu}B^{\mu\nu}$, where $X_\mu\nu$ ($B_\mu\nu$) is the Abelian hidden (SM) sector field strengh. 
The consequence is that not only are the neutral current couplings modified by the $Z$-$Z'$ mass mixing angle $\eta$, they contain an extra piece proportional to the fermion hypercharge due to the $U(1)$ mixing~\cite{CNW06}. In this scenario, the extra $Z'$ can be very light. In the case where $M_Z' \ll M_W$, $\delta_{LR}$ is dominated by the photon-$Z'$ interference: 
$\delta_{LR}\simeq(s_\eta^2 - 15.80 c_\eta^2 s_\eps^2 + 31.85 c_\eta s_\eta s_\eps)(M_Z^2/M_{Z'}^2)$, where $c$ ($s$) denotes cosine (sine), and $\eps$ characterizes the gauge kinetic mixing. Rewriting $s_\eps$ in terms of $s_\eta$ and $M_Z'$~\cite{CNW09}, the allowed parameter space is shown by the contour plot of $\delta_{LR}$ in Fig.~\ref{fig:hidZ}. Nota that there is no allowed parameter space for $\delta_{LR} > 0$, and so the sign of $\delta_{LR}$ can be used to tell whether the $Z'$ comes from a high or low scale.

\section{Summary}
High-precision polarized M\o{}ller scattering experiment will be a very useful complementary probe of NP at the several TeV order. When operate in tandem with the LHC, it can provide a clean determination of couplings of the new states discovered at the LHC -- e.g. extra $Z'$ gauge bosons or doubly charged scalars -- 
to electrons, as well as putting useful constraints on the mixing parameters in various scenarios. By measuring the sign of $\delta_{LR}$, it can also help determine the origin of these new states.

On its own, high-precision M\o{}ller scattering can probe low scale hidden sector, which is inaccessible to high energy colliders like the LHC. It can measure the coupling of the hidden sector to the SM through mixing, and constrain the degree of this mixing. This will provide valuable inputs to for further dark matter model building employing a hidden sector.

\end{document}